\documentclass[reprint,prl]{revtex4-1}
\usepackage{bm}
\usepackage{graphicx}
\usepackage[usenames]{color}
\usepackage{amsmath,amsfonts,amssymb}
\usepackage{hyperref}

\begin{document}

\title{Hannay Angle: Yet Another Symmetry Protected Topological Order
Parameter in Classical Mechanics}

\date{\today}
\author{Toshikaze Kariyado}\email{kariyado@rhodia.ph.tsukuba.ac.jp}
\author{Yasuhiro Hatsugai}\email{hatsugai@rhodia.ph.tsukuba.ac.jp}
\affiliation{Division of Physics, Faculty of Pure and Applied Sciences,
University of Tsukuba, Tsukuba, Ibaraki 305-8571, Japan}

  \begin{abstract}
   Topological way of thinking now goes beyond conventional solid
   materials, and topological characterization of classical mechanical
   systems governed by Newton's equation of motion begins to attract
   much attention. To have a deeper insight on
   physical meaning of topological numbers in mechanical systems,
   we demonstrate the use of the Hannay angle, a classical counterpart
   of the Berry phase, as a symmetry protected topological order
   parameter. We first derive the Hannay angle using a
   canonical transformation that maps the Newton's equation to the
   Schr\"{o}dinger type equation. The Hannay
   angle is then used to characterize a simple spring-mass model
   topologically with a particular focus on the bulk-edge correspondence
   and new aspects of the symmetry in a classical system.
  \end{abstract}

\maketitle

A study of topological phenomena in solids has its origin in quantum
Hall systems \cite{PhysRevLett.45.494}. However, topological phenomena
are not limited to quantum
systems, but also found in classical systems such as photonic crystals
governed by the Maxwell equation
\cite{PhysRevLett.100.013905,PhysRevLett.100.013904}. Very recently,
classical mechanical
systems obeying the Newton's equation of motion begin to get much attention as a playground of topology
\cite{PhysRevLett.103.248101,Chen09092014,Kane:2014aa,Po:2014,wang:2015,nash:2015,huber:2015,1367-2630-17-7-073031}.
Although topology is an abstract mathematical concept,
physical observables are characterized by the topological quantities
and well described.
For instance, in the quantum Hall systems, a topological number, i.e.,
the Chern number defined using the Bloch wave functions
\cite{PhysRevLett.49.405}, has a direct relation to
the Hall conductance and the number of edge states
\cite{PhysRevLett.71.3697}. Another example is
the Berry phase, which describes the topological edge states of Dirac
fermions in solid state materials such as graphene
\cite{PhysRevLett.89.077002,PhysRevB.84.195452,PhysRevB.88.245126,PhysRevB.90.085132}. In
contrast to the Chern number, which is quantized by definition, the
Berry phase is quantized and topological only with
a help of some appropriate symmetry. It is the symmetry protection of
topological phases.
In any cases, nontrivial bulk as a topological phase implies
  existence of localized edge states if the system has boundaries, i.e.,
  there is the bulk-edge correspondence.
  This is surely quite important for topological description of
  classical systems, where we do not mostly have bulk topological observables such as the Hall
  conductance. Even though the classical
  system is topologically nontrivial, we may only access
  its toplogical character through the edge states. Direct access to the
  bulk topological observable in classical systems is an important issue
  to be addressed. 

As a first step toward better understanding of topological numbers in
classical systems, here we demonstrate the use
of the Hannay angle \cite{0305-4470-18-2-011},
which is a classical counterpart of the Berry phase, as a symmetry
protected topological order parameter. We first present a concise
formulation of the geometrical angle, i.e., the Hannay angle, in
mechanical system using a canonical transformation, such that the
Newton's equation is mapped to the Schr\"{o}dinger type equation.
Then, the solid relation between the
quantized Hannay angle by symmetry and the edge states is established by
numerically analyzing a simple spring-mass model. Interestingly, the
symmetry protected nature is observed as boundary condition
dependence of the edge modes. 

Let us begin with describing a geometrical phase in a classical system
using a simple example, 2D motion of a mass point (mass
$m=1$ is assumed for simplicity) in the
potential $V_\phi(x,y)=\alpha^2{}^t\!\bm{x}\hat{\Gamma}_\phi\bm{x}/2$ with 
\begin{equation}
 \hat{\Gamma}_\phi=
  \hat{1}+\Delta(1-\eta+\eta\cos\phi)\hat{\sigma}_3+\Delta\eta\sin\phi\hat{\sigma}_1, \label{eq:Gamma_phi}
\end{equation}
where $\phi$ has some time dependence and $\sigma_i$ is the Pauli
matrix. $V_\phi(x,y)$ is periodic in
$\phi$ with period $2\pi$. This problem is so simple, but captures
essence of the geometrical phase in
classical mechanics. For the fixed $\phi$, this system has two normal
modes with frequencies
$\omega^{\pm}_\phi=\alpha\sqrt{1\pm\Delta\sqrt{1-2\eta(1-\eta)(1-\cos\phi)}}$. Now,
assuming the time dependence of $\phi$ as $\Omega{t}$, we follow the
time evolution starting from the initial conditions $\bm{x}={}^t\!(0,1)$
and $\dot{\bm{x}}=\bm{0}$, which excites only the mode with
$\omega^-_0$. We assume slow evolution of $\phi$, i.e.,
$\Omega/\alpha\ll \Delta$, and then, the other mode with $\omega^+_\phi$
is expected to be unexcited during the time evolution. In order to
factor out rapid oscillation, which corresponds to the dynamical
contribution, new dynamical variables $\bm{x}_c={}^t\!(x_c,y_c)$ and
$\bm{x}_s={}^t\!(x_s,y_s)$ are introduced as
$\bm{x}=\bm{x}_c\cos\theta_t+\bm{x}_s\sin\theta_t$ with
$\theta_t=\int_0^t\!\omega^-_{\Omega{t'}}\mathrm{d}t'$, 
and the equation of motion is divided into the coefficients of
$\cos\theta_t$ and $\sin\theta_t$.

\begin{figure}[tbp]
 \begin{center}
  \includegraphics[scale=1.0]{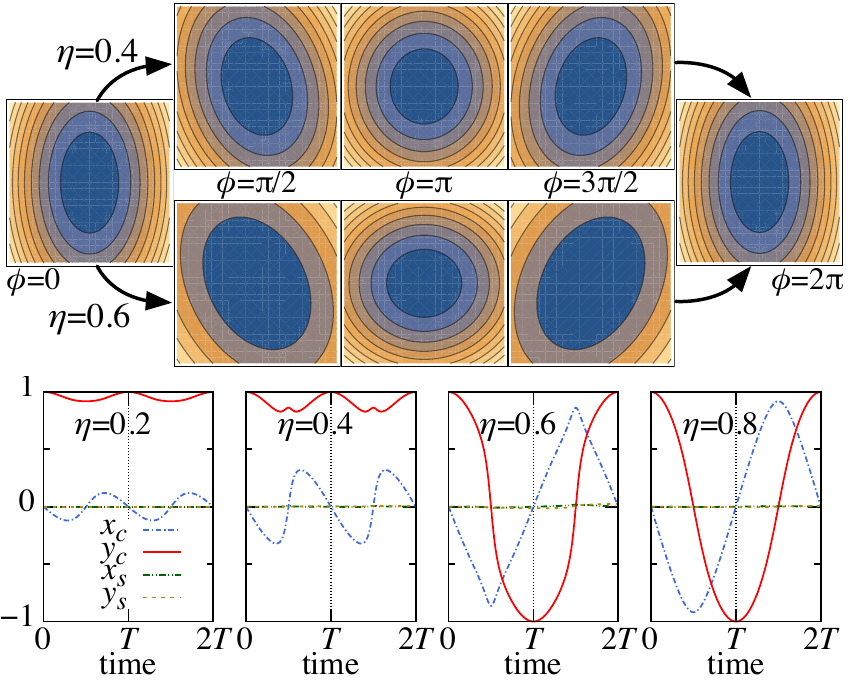}
  \caption{Upper panel: Contour plots of the potential $V_\phi(x,y)$ for
  $\eta=0.4$ and $0.6$. Lower panels: Time evolution of $\bm{x}_c$ and
  $\bm{x}_s$ with $\alpha=2\pi$, $\Delta=0.5$, and $T=1200$ for selected
  $\eta$.
  }\label{fig1}
 \end{center}
\end{figure}

The numerically obtained time
evolution of $\bm{x}_c$ and $\bm{x}_s$ is plotted in Fig.~\ref{fig1}. We
find that at $t=T$, when the potential gets back to the original form,
$\bm{x}_c(T)$ is same as $\bm{x}_c(0)$ for $\eta<0.5$ while
$\bm{x}_c(T)$ is minus of $\bm{x}_c(0)$ for $\eta>0.5$. Since the rapid
oscillation is already factored out, this means that the system acquires
extra $\pi$ phase other than the dynamical contribution for
$\eta>0.5$. The physical origin of this extra $\pi$ phase is actually
simple. Namely, $V_\phi(x,y)$ is constructed by superposing two
functions $v_1(x,y)=c_{11}x^2+c_{12}y^2$ and
$v_2(x,y)=c_{21}x^2+c_{22}y^2$ with appropriate $c_{ij}$, and the
evolution from $\phi=0$ to $\phi=2\pi$ is achieved by relatively
rotating $v_2(x,y)$ against $v_1(x,y)$ by 180$^\circ$, instead of
360$^\circ$. When $v_2(x,y)$ is dominant, 180$^\circ$ rotation
results in a flip of $\bm{x}_c$ after completion of one period, while
when $v_1(x,y)$ is dominant, such a flip does not take place. 

In order to interpret the extra $\pi$ phase as a geometrical phase, i.e.,
the Hannay angle, we
introduce action-angle variables \cite{Landau-Lifshitz,Arnold}. We start
with a Hamiltonian
$H=(\bm{p}'\cdot\bm{p}'+{}^t\!\bm{q}'\hat{\Gamma}_\lambda\bm{q}')/2$, where
$\bm{p}'$ and $\bm{q}'$ are canonical variables, and 
$\hat{\Gamma}_\lambda$ is a symmetric matrix describing the potential
that is parameterized by $\lambda$.
Using the idea of the Bargmann
variables \cite{0305-4470-26-15-029}, we introduce a canonical
transformation by a generation function
$F(\bm{q}',\bm{q})=-\mathrm{i}({}^t\!\bm{q}'-a_+{}^t\!\bm{q})\hat{\gamma}_\lambda(\bm{q}'-a_-\bm{q})/2$
with $a_\pm=\sqrt{2}\pm{1}$ and 
$\hat{\gamma}_\lambda=\sqrt{\hat{\Gamma}_\lambda}$, which 
leads to a transformed Hamiltonian
\begin{equation}
H=\mathrm{i}{}^t\!\bm{p}\hat{\gamma}_\lambda\bm{q}.\label{qp-hamilt}
\end{equation}
The ``offdiagonal'' form with $\sqrt{\hat{\Gamma}_\lambda}$ reminds us the
``supersymmetric'' treatment of the spring mass model proposed by Kane
and Lubensky \cite{Kane:2014aa}.
The equation of motion is now written as
\begin{equation}
 \dot{\bm{q}}=\partial{H}/\partial\bm{p}=\mathrm{i}\hat{\gamma}_\lambda\bm{q},
  \quad
  \dot{\bm{p}}=-\partial{H}/\partial\bm{q}=-\mathrm{i}\hat{\gamma}_\lambda\bm{p} \label{EoM_PQ}
\end{equation}
with general solutions
 $\bm{q}(t)=\mathrm{e}^{\mathrm{i}\hat{\gamma}_\lambda{t}}\bm{q}_0$ and 
 $\bm{p}(t)=\bm{p}_0\mathrm{e}^{-\mathrm{i}\hat{\gamma}_\lambda{t}}$.
Since Eq.~\eqref{EoM_PQ} is exactly like the Schr\"{o}dinger equation,
i.e., it only contains the first order derivative in time, it is
possible to give a ``quantum'' derivation of the Hannay angle. Namely,
we can regard $\hat{\gamma}_\lambda\bm{q}$ and $\mathrm{i}\bm{p}$ as a
ket $|\psi\rangle$ and a bra
$\langle\psi|$, respectively, and follow the time evolution in a quantum
fashion. (The variables $\bm{q}$ and $\bm{p}$ are
independent, but in order to keep $\bm{q}'$ and $\bm{p}'$ real, they
should fulfill the relation $(\mathrm{i}\bm{p})^*=\hat{\gamma}_\lambda\bm{q}$.)
Here, in order to emphasize the classical nature of the phenomenon, we derive
it in a purely classical fashion.

The action is defined as
\begin{equation}
 S(\bm{q},\bm{q}_0,\bm{p}_0,\lambda)=\int_{C(\bm{q},\bm{q}_0,\bm{p}_0,\lambda)}\bm{p}\cdot\mathrm{d}\bm{q},
\end{equation}
where $C(\bm{q},\bm{q}_0,\bm{p}_0,\lambda)$ is a path in the phase space
whose initial and final values of $\bm{q}(t)$ are $\bm{q}$ and
$\bm{q}_0$, and initial value of $\bm{p}(t)$ is $\bm{p}_0$,
respectively. Explicitly, we have
\begin{equation}
 S(\bm{q},\bm{q}_0,\bm{p}_0,\lambda)
  =\int\bm{p}\cdot\dot{\bm{q}}\mathrm{d}t
  =\sum_\alpha\log(Q_{\alpha}/Q_{0\alpha})P_{0\alpha}Q_{0\alpha}
\end{equation}
with $\bm{Q}=\hat{O}\bm{q}$ and $\bm{P}=\bm{p}\hat{O}^\dagger$ where
$\hat{O}$ is a matrix diagonalizing $\hat{\gamma}_\lambda$. Namely, 
$\hat{\gamma}_\lambda=\hat{O}^\dagger\hat{\omega}\hat{O}$ where $\hat{\omega}$
is diagonal with $\omega_\alpha$ as diagonal elements.
Now, the action variables $I_\alpha$ are introduced as
$I_\alpha=\mathrm{i}P_{0\alpha}Q_{0\alpha}$.
By definition, $2\pi I_\alpha$ represents increment of $S$ against
completion of one period of the normal mode with $\omega_\alpha$.
Using $\bm{I}$, we can express $S$ as a function of $\bm{q}$,
$\bm{q}_0$, $\bm{I}$, and $\lambda$. Then, we consider a canonical
transformation taking $S$ as a generation
function. Since $I_\alpha$ is an adiabatic invariant, $I_\alpha$ is an
ideal quantity as new momentum. The 
new coordinate conjugate to $I_\alpha$ is
\begin{equation}
 \varpi_\alpha=\frac{\partial{S}(\bm{q},\bm{q}_0,\bm{p}_0,\lambda)}{\partial{I}_\alpha}=-\mathrm{i}\log(Q_\alpha/Q_{0\alpha})=\omega_\alpha{t},
\end{equation}
which is named as the angle variable. From this, the motion of mass point
is obviously periodic in $\varpi_\alpha$ with period $2\pi$. Therefore,
difference in $\varpi_\alpha$ indicates difference in phase of
oscillating motion.

Next, we consider the case that $\lambda$ weakly depends on time. Then,
the generation function $S$ depends on time through the time dependence
of $\lambda$, and the Hamiltonian is transformed as
$H\rightarrow{}H+(\partial{S}/\partial\lambda)\dot{\lambda}$. Using the
transformed Hamiltonian, the equation of motion for $\varpi_\alpha$ becomes
$\dot{\varpi}_\alpha=\partial H/\partial I_\alpha
  +\partial^2 S(\bm{q},\bm{I},\lambda)/(\partial{I}_\alpha\partial\lambda)
  \dot{\lambda}$
and we formally obtain
 \begin{equation}
  \varpi_\alpha=\int\frac{\partial H}{\partial I_\alpha}\mathrm{d}t
  +\int\frac{\partial^2 S(\bm{q},\bm{I},\lambda)}{\partial{I}_\alpha\partial\lambda}\frac{\mathrm{d}\lambda}{\mathrm{d}t}\mathrm{d}t.
 \end{equation}
 The first term represents the dynamical contribution, whereas the
 second term represents the geometrical contribution, which is the
 Hannay angle $\theta^{(H)}_\alpha$.
Following the Berry's argument \cite{0305-4470-18-1-012},
$\theta^{(H)}_\alpha$ is evaluated as 
\begin{equation}
  \theta^{(H)}_\alpha
   =-\frac{\partial}{\partial{I}_\alpha}\oint\bm{p}\cdot\frac{\partial\bm{q}}{\partial\lambda}\mathrm{d}\lambda,
\end{equation}
for the case that $\lambda$ gets back to the original value after
evolution. 
In practice, we should average out rapid oscillations to have a
convenient formula. The best method for the averaging is applying 
integration over $\varpi_\alpha$ in the range $[0,2\pi]$
\cite{0305-4470-18-2-011}. Here, we
focus on the situation that only one normal mode is excited.
In that case, provided that $\alpha=\tilde{\alpha}$ corresponds to the
excited mode, $I_\alpha$ is finite only for $\alpha=\tilde{\alpha}$, 
and it is reasonable to take integration only over
$\varpi_{\tilde{\alpha}}$. We finally obtain
\begin{equation}
\begin{split}
  \theta^{(H)}_{\tilde{\alpha}}&=-\frac{\partial}{\partial{I}_{\tilde{\alpha}}}
  \oint\biggl(
 \int_0^{2\pi}\negthickspace
 \frac{\mathrm{d}\varpi_{\tilde{\alpha}}}{2\pi}
  \bm{p}(\bm{\varpi},\bm{I},\lambda)\cdot\frac{\partial\bm{q}(\bm{\varpi},\bm{I},\lambda)}{\partial\lambda}\biggr)\mathrm{d}\lambda \\
 &=-\frac{\partial}{\partial{I}_{\tilde{\alpha}}}
 \oint\biggl(
 \int_0^{2\pi}\negthickspace
 \frac{\mathrm{d}\varpi_{\tilde{\alpha}}}{2\pi}\sum_i
 P_{\tilde{\alpha}}{O}_{\tilde{\alpha}i}\partial_\lambda{(\hat{O}^\dagger)}_{i\tilde{\alpha}}Q_{\tilde{\alpha}}\biggr)\mathrm{d}\lambda \\
 &=\mathrm{i}\oint(\hat{O}\partial_\lambda\hat{O}^\dagger)_{\tilde{\alpha}\tilde{\alpha}}\mathrm{d}\lambda.
\end{split}\label{Hannay_PQ}
\end{equation}
If we regard $O_{\alpha{i}}$ as a wave function, Eq.~\eqref{Hannay_PQ}
is exactly the Berry phase.

Now, it is possible to give two interpretations for the flip of $\bm{x}$
at $t=T$ for $\eta>0.5$ in Fig.~\ref{fig1}. First, since
$\hat{\Gamma}_\phi$ is a symmetric matrix, $\hat{O}_{\lambda=\phi}$ in
Eq.~\eqref{Hannay_PQ} can be a real matrix, which results in
$\theta^{(H)}=0$. However, in this case, it is impossible to make
$\hat{O}_{\phi}$ continuous {\it and} periodic in $\phi$ with period
$2\pi$. Namely, if we force $\hat{O}_\phi$ to be continuous in $\phi$,
we have $\hat{O}_{2\pi}=-\hat{O}_0$, which explains the flip of
$\bm{x}$. On the other hand, it is possible to amend $\hat{O}_\phi$ so
that it becomes continuous and periodic by introducing a complex phase
factor, typically $\mathrm{e}^{\mathrm{i}\phi/2}$. Such a phase factor
leads to $\hat{O}_{2\pi}=\hat{O}_0$, but then, Eq.~\eqref{Hannay_PQ}
gives an extra phase factor of $\pi$, which explains the flip of
$\bm{x}$. Possibility of two kinds of interpretations corresponds to the
gauge degrees of freedom in quantum cases.

This argument indicates that $\theta^{(H)}$ should
be $0$ or $\pi$, since $\hat{O}_{2\pi}=\pm\hat{O}_0$ is the only
possibility with the gauge choice in which $\hat{O}_\phi$ becomes
real. However, $\theta^{(H)}$ deviates from $0$ or $\pi$ if the
starting Hamiltonian is changed to break the time reversal symmetry. As
an example, let us consider the situation that the system is rotated
with angular frequency $\Omega_R$. Then, the Hamiltonian written in the
rotating frame is obtained by replacing $\bm{p}'$ by
$\bm{p}'+\mathrm{i}\Omega_R\hat{\sigma}_y\bm{q}'$. Again, we follow the
time evolution of $\bm{x}_c$ and $\bm{x}_s$ introduced as
$\bm{q}'=\bm{x}_c\cos\theta_t+\bm{x}_s\sin\theta_t$, but with modified
$\omega^\pm_\phi=\alpha\sqrt{1+\beta^2\pm\sqrt{\Delta^2(1-2\eta(1-\eta)(1-\cos\phi))+4\beta^2}}$
where $\beta=\Omega_R/\alpha$. We use the initial condition
$\bm{x}={}^t\!(0,1)$ and $\dot{\bm{x}}={}^t\!(v_0,0)$ with
$v_0=-2\beta(1+\beta^2-\sqrt{\Delta^2+4\beta^2})/(\Delta-2\beta^2+\sqrt{\Delta^2+4\beta^2})$
in order to excite only the mode for $\omega^-_0$.

\begin{figure}[bt]
 \begin{center}
  \includegraphics[scale=1.0]{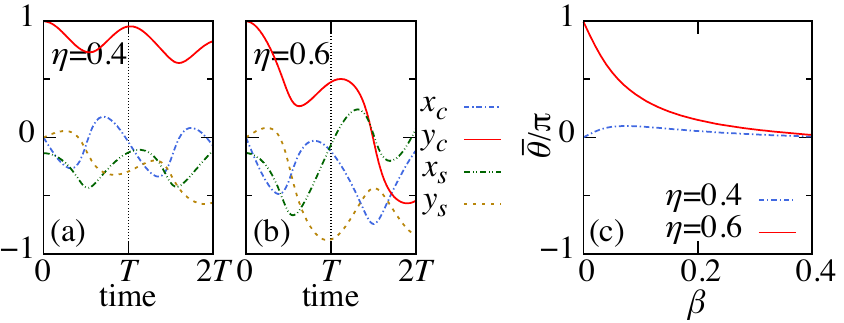}
  \caption{(a,b) Time evolution of $\bm{x}_c$ and $\bm{x}_s$ for
  $\eta=0.4$ and $0.6$ for $\beta=0.1$. (c) $\beta$ dependence of
  $\bar{\theta}$ for $\eta=0.4$ and $0.6$. The parameters $\alpha=2\pi$, $\Delta=0.5$, and $T=1200$ are used in (a)-(c).}\label{fig2}
 \end{center}
\end{figure}
The numerically
obtained ($\bm{x}_c$, $\bm{x}_s$) for $\eta=0.4$ and $\eta=0.6$ are
plotted in Figs.~\ref{fig2}(a) and \ref{fig2}(b). In contrast to
Fig.~\ref{fig1} where $x_s$ and $y_s$
are negligible for all time, $x_s$ and $y_s$ have significant
time dependence. Furthermore, we have $\bm{x}_c(T)\neq\pm\bm{x}_c(0)$, which
indicates the extra phase factor other than $0$ or $\pi$. With our
choice of the initial condition, the extra phase factor $\bar{\theta}$
corresponding to $\theta^{(H)}$
is evaluated as $\bar{\theta}=\arccos{y_c}(T)$, and plotted in Fig.~\ref{fig2} as
a function of $\beta=\Omega_R/\alpha$. We see that $\bar{\theta}$
clearly deviates
from $0$ and $\pi$ for finite $\Omega_R$. Now, it is desirable to find
out a transformation mapping the rotated Hamiltonian to the Hamiltonian
like Eq.~\eqref{qp-hamilt}. However, a naive extension of our
transformation by $F(\bm{q}',\bm{q})$ is only applicable in the case that
$[\sigma_y,\hat{\Gamma}_\phi]=0$, which limits us to the trivial case with
$\hat{\Gamma}_\phi\propto\hat{1}$.
A possible analytical
formulation for the rotated case is a future problem. 

\begin{figure}[tbp]
 \begin{center}
  \includegraphics[scale=1.0]{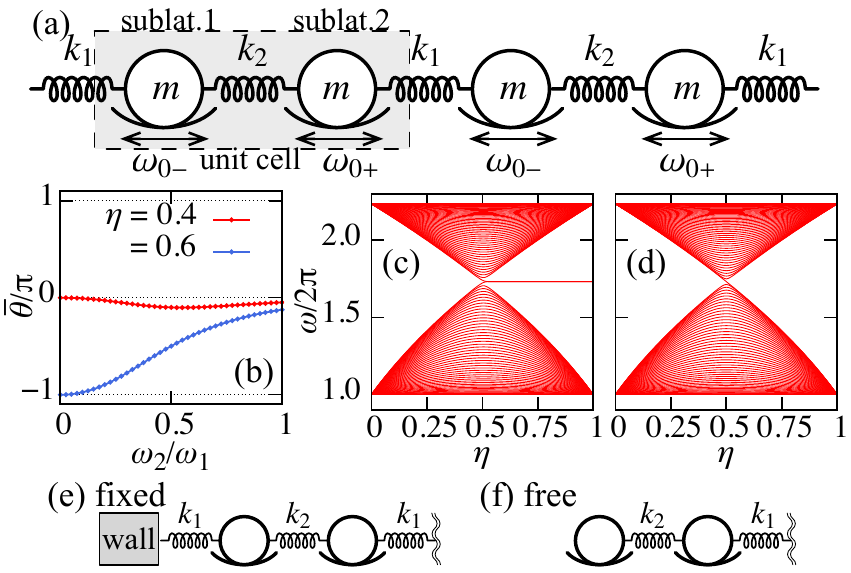}
  \caption{(a) Schematic picture of the ``dimerized''
  spring-mass model. An employed unit cell convention is also
  shown. (b) $\omega_2$ dependence of $\bar{\theta}$ in the spring-mass
  model for
  $\eta=0.4$ and $0.6$ with $\omega_0=\omega_1=2\pi$ and $T=800$. (c,d)
  Frequency spectra for the finite size spring-mass model with the (c)
  fixed and (d) free boundary condition. (e,f) The fixed and free boundary
  conditions.}\label{fig3}
 \end{center}
\end{figure}

For the demonstration of the usage of the Hannay angle as a symmetry
protected topological order parameter, we introduce a one-dimensional
``dimerized'' spring-mass model schematically depicted in
Fig.~\ref{fig3}(a). 
The involved parameters are mass of the mass points $m=1$, frequencies
associated with sublattice dependent harmonic potential $\omega_{0\pm}$,
and spring constants $k_1$ and $k_2$. For later
convenience, we introduce parameters $\omega_0$, $\omega_1$,
$\omega_2$, $\eta$ as $\omega_{0\pm}^2=\omega_0^2\pm\omega_2^2$,
$k_1=\eta\omega_1^2$ and $k_2=(1-\eta)\omega_1^2$. We limit the motion
of mass points along the 1D chain, then, the dynamical
variable for this system is deviation of each mass point from the
equilibrium position, which is written as $x_{na}$ for the mass point at
$n$th unit cell and sublattice $a$. Assuming the periodic boundary
condition and introducing new variables $f_{ka}$ as 
$x_{na}=\sum_{k}\mathrm{e}^{-\mathrm{i}nk}f_{ka}$, the
equation to be solved becomes $\ddot{\bm{f}}_k=-\hat{H}_k\bm{f}_{k}$
where $\bm{f}_k={}^t\!(f_{k1},f_{k2})$, and 
\begin{equation}
 \hat{H}_k=(\omega_0^2+\omega_1^2)\hat{1}-\omega_1^2\textrm{Re}g_k\hat{\sigma}_1-\omega_1^2\textrm{Im}g_k\hat{\sigma}_2+\omega_2^2\hat{\sigma}_3 \label{eq:Hk}
\end{equation}
with $g_k=(1-\eta)+\eta\mathrm{e}^{\mathrm{i}k}$. Now, $\hat{H}_k$
contains complex elements, but it is straightforward to extend the
former arguments for the case that $\hat{\Gamma}$ is real symmetric \cite{comment1}. For fixed $k$,
there exist two eigenmodes with frequencies
$\omega_{\pm}(k)=\sqrt{\omega_0^2+\omega_1^2\pm\sqrt{\omega_2^2+\omega_1^2|g(k)|^2}}$. 
When $\omega_2=0$, $\hat{H}_k$ does not contain $\hat{\sigma}_3$ and
then, we say that the system has a ``chiral symmetry'' since
$\hat{H}_k-(\omega_0^2+\omega_1^2)\hat{1}$ anticommutes with
$\hat{\sigma}_3$. Although it is not the chiral symmetry in a strict
sense because of the constant diagonal part, this symmetry is important
in topological characterization. Furthermore, for $\omega_2=0$, it is
possible to map Eq.~\eqref{eq:Hk} to Eq.~\eqref{eq:Gamma_phi} by a
parameter independent transformation such that
$\hat{\sigma}_1\rightarrow\hat{\sigma}_3$ and $\hat{\sigma}_2\rightarrow\hat{\sigma}_1$, which
indicates that the two problems are essentially identical to each other.

Next, we follow the time evolution regarding $k$ as a controllable
parameter driven as $k=\Omega{t}$. More specifically, 
we solve
\begin{equation}
 \frac{\mathrm{d}^2\bm{f}(t)}{\mathrm{d}t^2}=-\hat{H}_{\Omega
  t}\bm{f}(t),
\end{equation}
with the initial condition
$\bm{f}(0)=\bm{f}_0={}^t\!(1/\sqrt{2},1/\sqrt{2})$ and
$\dot{\bm{f}}(0)=0$, which excites only the lower branch with $\omega_-(0)$ at $t=0$. 
In order to factor out the rapid oscillation, i.e., the dynamical
contribution to $\varpi_\alpha$, new variables $\bm{u}_\pm(t)$ are
introduced as
$\bm{f}(t)=\mathrm{e}^{\mathrm{i}\theta_t}\bm{u}_+(t)+\mathrm{e}^{-\mathrm{i}\theta_t}\bm{u}_-(t)$ with
$\theta_t=\int^t\!_0\omega^-_{\Omega\tau}\mathrm{d}\tau$,
and we solve equations for $\bm{u}_\pm(t)$. The initial condition
$(\bm{f}(0),\dot{\bm{f}}(0))=(\bm{f}_0,0)$ is satisfied by the choice of
$\bm{u}_{\pm}(0)=\bm{f}_0/2$ and
$\dot{\bm{u}}_\pm(0)=0$. Figure~\ref{fig2}(b) shows the $\omega_2$
dependence of
$\bar{\theta}=\mathrm{arg}(\bm{u}_+^\dagger(0)\cdot\bm{u}_+(T))$ for
several $\eta$. Since the dynamical contribution is factored out,
$\bar{\theta}$ captures the geometrical contribution. For $\omega_2=0$,
since $\hat{H}_k$ can be mapped to $\hat{\Gamma}_\phi$, $\bar{\theta}=0$
for $\eta<0.5$, while $\bar{\theta}=\pi$ for $\eta>0.5$. On the other
hand, for the finite $\omega_2$, which breaks the ``chiral symmetry'',
$\bar{\theta}$ deviates from $0$ or $\pi$. In short, the quantization of
the Hannay angle into $0$ or $\pi$ is protected by the ``chiral
symmetry''. Due to Eq.~\eqref{Hannay_PQ}, it is natural that the
quantization condition for the Hannay angle is the same as for the Berry
phase. 

The bulk-edge correspondence is useful in connecting the quantized
Hannay angle and the topological character of
the system in terms of edge states. In Fig.~\ref{fig3}(c), the frequency
spectrum for the finite length system having the fixed boundary with
$\omega_2=0$ is plotted as a function of $\eta$. There is a clear
signature of in-gap edge state distinct
from the bulk contribution for $\eta>0.5$ where $\bar{\theta}=\pi$,
while there is no sign of edge state for $\eta<0.5$ where
$\bar{\theta}=0$. Therefore, the quantized Hannay angle
is used to detect the topological transition characterized by
the appearance of localized edge modes. Here, we should be careful on the unit
cell convention to establish the relation between the Hannay angle
and edge states, since the different unit
cell conventions lead to the modified Hamiltonian, which affects the
Hannay angle \cite{PhysRevB.88.245126}. In specific, the convention such
that the boundary is in between the two neighboring unit cells should be
employed. 

Figure~\ref{fig3}(b) shows the frequency spectrum obtained with the free
boundary condition. In contrast to the fixed boundary case, there is no
sign of in-gap edge state irrespective of the value of $\eta$.
As discussed in the similar situation \cite{mech_graphene}, this
contrast is actually the manifestation of the symmetry protection of the
topological states. Recall that the
quantization of the Hannay angle is protected by the ``chiral
symmetry''. For the fixed boundary condition, the ``chiral symmetry'' is
preserved even after the boundary is introduced, since the diagonal
elements of $\hat{H}$ are uniform even with the boundary. 
On the other
hand, the free boundary breaks the ``chiral symmetry'', since the
diagonal elements of $\hat{H}$ becomes
nonuniform due to the absence of a spring at the free boundary
\cite{mech_graphene}. Then, the edge mode is not necessarily
protected because of the symmetry breaking, even if the bulk parts share
the same topological character for the two kinds of boundary conditions. 

To summarize, we derive a concise formula for the geometrical
angle, i.e., the Hannay angle, in mechanical system by making use of the
canonical transformation that maps the Newtonian equation of motion to
the Schr\"{o}dinger type equation. Then, the use
of the Hannay angle
as a symmetry protected topological order parameter is demonstrated
using a simple spring-mass model. The importance of the symmetry in
topological characterization is pointed out by analyzing the boundary
condition dependence of the edge modes.
It reveals yet another role
of the symmetry in mechanical systems in relation to the bulk-edge
correspondence.

\section*{Acknowledgements}
\begin{acknowledgments}
 The work is partly supported by Grants-in-Aid for Scientific Research
 No. 26247064 from JSPS and No. 25107005 from MEXT.
\end{acknowledgments}

 \appendix
 \section{Appendix: The Case with Hermitian $\hat{\Gamma}$}
Let us start with the Hamiltonian
\begin{equation}
 H=p'_i\bar{p}'_i+\bar{q}'_i\Gamma_{ij}q'_j.
\end{equation}
where $\hat{\Gamma}$ is an Hermitian matrix. 
Now, we consider a canonical transformation using a generation function
\begin{multline}
 F(q'_i,\bar{q}'_i,q_i,\bar{q}_i)\\
 =-\mathrm{i}\bar{q}'_i\gamma_{ij}q'_j
  +\mathrm{i}\sqrt{2}\bar{q}'_i\gamma_{ij}q_j
  +\mathrm{i}\sqrt{2}\bar{q}_i\gamma_{ij}q'_j
  -\mathrm{i}\bar{q}_i\gamma_{ij}q_j.
\end{multline}
Here $\hat{\gamma}=\sqrt{\hat{\Gamma}}$ and $\hat{\gamma}$ is hermitian
under the condition that $\hat{\Gamma}$ is positive semidefinite. In the
following, we assume this condition is fulfilled. 
Then, we have
\begin{align}
 p'_i&=\frac{\partial F}{\partial q'_i}
 =-\mathrm{i}\bar{q}'_j\gamma_{ji}+\mathrm{i}\sqrt{2}\bar{q}_j\gamma_{ji},\\
 \bar{p}'_i&=\frac{\partial F}{\partial \bar{q}'_i}
 =-\mathrm{i}\gamma_{ij}q'_j
 +\mathrm{i}\sqrt{2}\gamma_{ij}q_j,\\
 p_i&=-\frac{\partial F}{\partial q_i}
 =-\mathrm{i}\sqrt{2}\bar{q}'_{j}\gamma_{ji}+\mathrm{i}\bar{q}_j\gamma_{ji},\\
 \bar{p}_i&=-\frac{\partial F}{\partial \bar{q}_i}
 =-\mathrm{i}\sqrt{2}\gamma_{ij}q'_j+\mathrm{i}\gamma_{ij}q_j,
\end{align}
which gives us
\begin{align}
 \mathrm{i}p'_i&=\frac{1}{\sqrt{2}}(\mathrm{i}p_i-\bar{q}_j\gamma_{ji}),&
 \mathrm{i}\bar{p}'_i&=\frac{1}{\sqrt{2}}(\mathrm{i}\bar{p}_i-\gamma_{ij}q_j),\\
 \gamma_{ij}q'_j&=\frac{1}{\sqrt{2}}(\gamma_{ij}q_j+\mathrm{i}\bar{p}_i),&
 \bar{q}'_{j}\gamma_{ji}&=\frac{1}{\sqrt{2}}(\bar{q}_j\gamma_{ji}+\mathrm{i}p_i).
\end{align}
and
\begin{align}
 \mathrm{i}p_i&=\frac{1}{\sqrt{2}}(\mathrm{i}p'_i+\bar{q}'_j\gamma_{ji}),&
 \mathrm{i}\bar{p}_i&=\frac{1}{\sqrt{2}}(\mathrm{i}\bar{p}'_i+\gamma_{ij}q'_j),\\
 \gamma_{ij}q_j&=\frac{1}{\sqrt{2}}(\gamma_{ij}q'_j-\mathrm{i}\bar{p}'_i),&
 \bar{q}_{j}\gamma_{ji}&=\frac{1}{\sqrt{2}}(\bar{q}'_j\gamma_{ji}-\mathrm{i}p'_i).
\end{align}
The Hamiltonian is transformed as
\begin{equation}
\begin{split}
  H=&-\frac{1}{2}\bigl(\mathrm{i}p_l-\bar{q}_i\gamma_{il}\bigr)
  \bigl(\mathrm{i}\bar{p}_l-\gamma_{lj}q_j\bigr)\\
 &\quad\quad
 +\frac{1}{2}\bigl(\mathrm{i}p_l+\bar{q}_i\gamma_{il}\bigr)
 \bigl(\mathrm{i}\bar{p}_l+\gamma_{lj}q_j\bigr)\\
 =&\mathrm{i}p_i\gamma_{ij}q_j+\mathrm{i}\bar{q}_i\gamma_{ij}\bar{p}_j.
\end{split}
\end{equation}
Then,the equations of motion are given as
\begin{align}
 \dot{q}_i&=\mathrm{i}\gamma_{ij}q_j,&
 \dot{\bar{q}}_i&=\mathrm{i}\bar{q}_j\gamma_{ji},\\
 \dot{p}_i&=-\mathrm{i}p_j\gamma_{ji},&
 \dot{\bar{p}}_i&=-\mathrm{i}\gamma_{ij}\bar{p}_j.
\end{align}
If we choose an initial condition such that $(q'_i)^*=\bar{q}'_i$ and
$(p'_i)^*=\bar{p}'_i$, it is mapped to an initial condition such that
$(\mathrm{i}p_i)^*=\gamma_{ij}q_j$ and
$(\mathrm{i}\bar{p})_i^*=\bar{q}_j\gamma_{ji}$. Since the equation of
motion implies that $\mathrm{i}p_i$ and $\gamma_{ij}q_j$, and
$\mathrm{i}\bar{p}_i$ and $\bar{q}_j\gamma_{ji}$ obey
conjugate equations, respectively, such relationships are preserved in
time evolution. On the other hand, the relations
$(\mathrm{i}p_i)^*=\gamma_{ij}q_j$ and
$(\mathrm{i}\bar{p}_i)^*=\bar{q}_j\gamma_{ji}$ imply $(q'_i)^*=\bar{q}'_i$ and
$(p'_i)^*=\bar{p}'_i$.

\end{document}